\documentclass[conference]{IEEEtran}
\IEEEoverridecommandlockouts
\usepackage{cite}
\usepackage{amsmath,amssymb,amsfonts}
\usepackage{algorithmic}
\usepackage[dvipdfmx]{graphicx}
\usepackage{textcomp}
\usepackage{xcolor}
\def\BibTeX{{\rm B\kern-.05em{\sc i\kern-.025em b}\kern-.08em
    T\kern-.1667em\lower.7ex\hbox{E}\kern-.125emX}}

\usepackage{bm,cite}
\usepackage[caption=false,font=footnotesize]{subfig}
\usepackage{algorithm,algorithmic}

\newcommand{\minimize}{\mathop{\rm minimize}\limits}

\begin{document}

\title{Sound Field Reproduction With Weighted Mode Matching and Infinite-Dimensional Harmonic Analysis: An Experimental Evaluation\\
\thanks{This work was supported by JST PRESTO Grant Number JPMJPR18J4.}
}

\author{\IEEEauthorblockN{Shoichi Koyama}
\IEEEauthorblockA{\textit{Graduate School of Information Science and Technology} \\
\textit{The University of Tokyo}\\
Tokyo, Japan \\
koyama.shoichi@ieee.org}
\and
\IEEEauthorblockN{Keisuke Kimura}
\IEEEauthorblockA{\textit{Graduate School of Information Science and Technology} \\
\textit{The University of Tokyo}\\
Tokyo, Japan \\
tree-village-ut@g.ecc.u-tokyo.ac.jp}
\and
\IEEEauthorblockN{Natsuki Ueno}
\IEEEauthorblockA{\textit{Graduate School of Information Science and Technology} \\
\textit{The University of Tokyo}\\
Tokyo, Japan \\
natsuki.ueno@ieee.org}
}

\maketitle

\begin{abstract}
Sound field reproduction methods based on numerical optimization, which aim to minimize the error between synthesized and desired sound fields, are useful in many practical scenarios because of their flexibility in the array geometry of loudspeakers. However, the reproduction performance of these methods in a practical environment has not been sufficiently investigated. We evaluate weighted mode matching, which is a sound field reproduction method based on the spherical wavefunction expansion of the sound field, in comparison with conventional pressure matching. We also introduce a method of infinite-dimensional harmonic analysis for estimating the expansion coefficients of the sound field from microphone measurements. Experimental results indicated that weighted mode matching using the expansion coefficients of the transfer functions estimated by the infinite-dimensional harmonic analysis outperforms conventional pressure matching, especially when the number of microphones is small. 
%Weighted mode matching is a sound field reproduction method based on the synthesis of harmonic coefficients of the desired sound field with multiple loudspeakers. Whereas analytical methods formulated in a continuous system require a simple array geometry, the weighted mode matching allows using arbitrarily placed loudspeakers. A weighting factor of each harmonic coefficient is determined depending on the setting of a target reproduction region. Thus, empirical truncation of the harmonic coefficients as used in the conventional mode matching is not necessary. We combine a sound field estimation method based on harmonic analysis of infinite order to incorporate harmonic coefficients of loudspeaker transfer functions estimated by distributed microphones. We report on experimental results in a practical environment for evaluating our proposed method. 
\end{abstract}


\section{Introduction}

The aim of sound field reproduction is to synthesize spatial sound by using multiple loudspeakers (or secondary sources), which has various applications such as virtual/augmented reality audio, generation of multiple sound zones for personal audio, and noise cancellation in a spatial region. In some applications, the sound field to be reproduced, i.e., the desired sound field, is captured by multiple microphones, which is called sound field recording. 

There are two major categories of sound field reproduction methods. One category is composed of analytical methods based on the boundary integral representations derived from the Helmholtz equation, such as wave field synthesis and higher-order ambisonics~\cite{Berkhout:JASA1980,Spors:AES124conv,Poletti:J_AES_2005,Ahrens:Acustica2008,Wu:IEEE_J_ASLP2009,Koyama:IEEE_J_ASLP2013,Koyama:JASA_J_2016}. The other category is composed of numerical methods based on minimization of the error between synthesized and desired sound fields inside a target region, such as pressure matching and mode matching~\cite{Poletti:J_AES_2005,Miyoshi:IEEE_J_ASSP_1988,Kirkeby:JASA_J_1996,Daniel:AES114conv,Betlehem:JASA_J_2005,Ueno:IEEE_ACM_J_ASLP2019}. Many analytical methods require the array geometry of the loudspeakers to have simple shape, such as a sphere, plane, circle, or line, and driving signals are obtained from a discrete approximation of the integral equation. On the other hand, in the numerical methods, the loudspeaker placement can be flexible, and driving signals are generally derived as a closed-form least-squares solution. 

Since the region in which the loudspeakers can be placed is limited in practical situations, a flexible loudspeaker array geometry in the numerical methods is preferable. We have recently proposed a numerical method of sound field reproduction, which is referred to as \textit{weighted mode matching}~\cite{Ueno:IEEE_ACM_J_ASLP2019}. This method is a generalization of mode matching, where both methods are based on the expansion of the sound field using spherical wavefunctions. The weighting factors for each expansion coefficient are derived on the basis of the setting of the target region and expected regional accuracy, which means that the empirical truncation of expansion order used in standard mode matching is unnecessary. 

In (weighted) mode matching, the transfer functions of the loudspeakers and the desired sound field must be represented by spherical wavefunction expansions. When it is difficult to model the transfer functions as simple analytical functions such as point sources, e.g., owing to reverberation, they have to be measured using microphones in advance. In sound field recording, the desired sound field is obtained by capturing a sound field using microphones. In both cases, the expansion coefficients of the spherical wavefunctions of the sound field must be estimated from microphone measurements. A spherical microphone array is typically used for this estimation~\cite{Meyer:ICASSP_2002,Abhayapala:ICASSP_2002,Park:JASA_J_2005,Rafaely:IEEE_J_ASLP_2005,Poletti:J_AES_2005,Koyama:JASA_J_2016}; however, to capture the sound field in a large region, the spherical microphone array should also be large, which is infeasible in many practical situations. It will be efficient to place multiple microphones or small microphone arrays over the target region rather than a single large spherical microphone array. We have recently proposed a method of estimating expansion coefficients on the basis of the harmonic analysis of infinite order~\cite{Ueno:IEEE_SPL2018,Ueno:IEEE_J_SP_2021}. Since this method is applicable to arbitrarily placed microphones, it will be useful to develop a flexible and scalable recording system.

In this study, we evaluate the reproduction performance of weighted mode matching using real data measured in a practical environment, in comparison with that of the pressure matching method. To estimate the expansion coefficients of the transfer functions of the loudspeakers, we applied infinite-dimensional harmonic analysis. In the experiment, the impulse response dataset MeshRIR recently published by the authors is used~\cite{Koyama:WASPAA2021}. The codes for reproducing the results are available in the example codes of MeshRIR.

\section{Problem Formulation}

\begin{figure}[t]
 \centerline{\includegraphics[width=0.6\columnwidth]{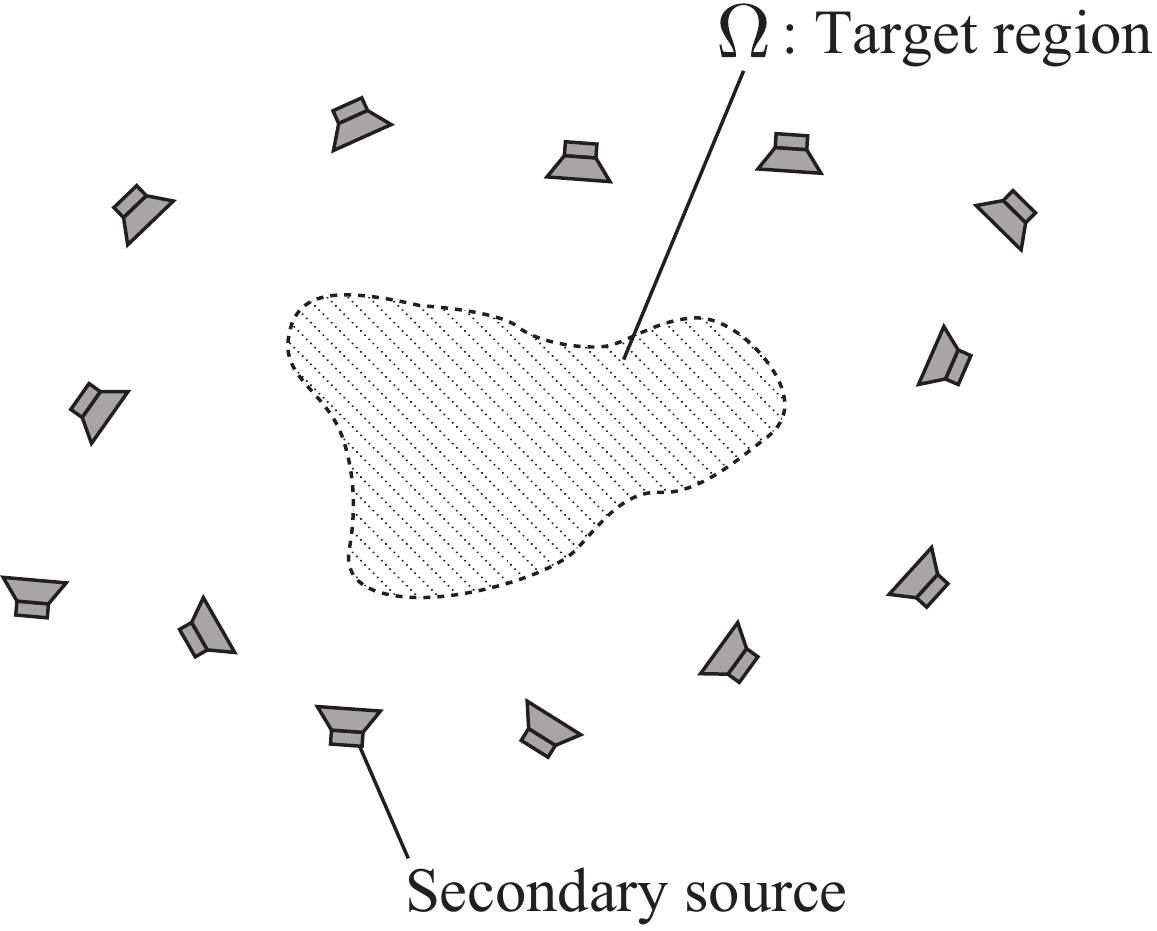}}
 \caption{Desired sound field is reproduced inside the target region $\Omega$ by using multiple secondary sources.}
\label{fig:sfc}
\end{figure}

We consider the synthesis of a given desired sound field in the interior region of a target $\Omega$ using multiple secondary sources. Suppose that $L$ secondary sources are arbitrarily placed in the region outside $\Omega$, i.e., $\mathbb{R}^3 \backslash \Omega$, as shown in Fig.~\ref{fig:sfc}. The desired sound field at the angular frequency $\omega$ is denoted by $u_{\mathrm{des}}(\bm{r},\omega)$, where $\bm{r}$ is the position vector in $\Omega$, i.e., $\bm{r}\in\Omega$. The sound field $u_{\mathrm{syn}}(\bm{r},\omega)$ at $\bm{r} \in D$ and $\omega$ synthesized using the secondary sources is represented as
\begin{align}
 u_{\mathrm{syn}}(\bm{r},\omega) = \sum_{l=1}^L d_l(\omega) g_l(\bm{r},\omega),
\label{eq:syn}
\end{align}
where $d_l(\omega)$ is the driving signal, and $g_l(\bm{r},\omega)$ is the transfer function from the $l$th secondary source to the position $\bm{r}$. Here, $g_l(\bm{r},\omega)$ ($l\in\{1,\ldots,L\}$) are assumed to be known by measuring or modeling them in advance. The goal of sound field reproduction is to obtain $d_l(\omega)$ of the $L$ secondary sources so that $u_{\mathrm{syn}}(\bm{r},\omega)$ coincides with $u_{\mathrm{des}}(\bm{r},\omega)$ inside $\Omega$. The argument $\omega$ is hereafter omitted for notational simplicity.

The driving signal $d_l$ for $l\in\{1,\ldots,L\}$ can be determined by minimizing the objective function
\begin{align}
 J %&= \int_{\bm{r}\in\Omega} \left| u_{\mathrm{syn}}(\bm{r},\omega) - u_{\mathrm{des}}(\bm{r}) \right|^2 \mathrm{d}\bm{r} \notag\\
   &= \int_{\Omega} \left| \sum_{l=1}^L d_l g_l(\bm{r}) - u_{\mathrm{des}}(\bm{r}) \right|^2 \mathrm{d}\bm{r} \notag\\
   &= \int_{\Omega} \left| \bm{g}(\bm{r})^{\mathsf{T}}\bm{d} - u_{\mathrm{des}}(\bm{r}) \right|^2 \mathrm{d}\bm{r},
\label{eq:obj}
\end{align}
where $\bm{g}(\bm{r})=[g_1(\bm{r}),\ldots,g_L(\bm{r})]^{\mathsf{T}}\in\mathbb{C}^L$ and $\bm{d}=[d_1,\ldots,d_L]^{\mathsf{T}}\in\mathbb{C}^L$ are the vectors of the transfer functions and driving signals, respectively. The minimization problem of Eq.~\eqref{eq:obj} is to minimize the mean square error of the reproduction in $\Omega$. In general, this problem cannot be analytically solved owing to the regional integral over $\Omega$. To incorporate the expected regional accuracy, a weighting function $\rho(\bm{r})$ ($\bm{r}\in\Omega$) is sometimes used as~\cite{Ueno:IEEE_ACM_J_ASLP2019}
\begin{align}
 J_{\rho} &= \int_{\bm{r}\in\Omega} \rho(\bm{r}) \left| \bm{g}(\bm{r})^{\mathsf{T}}\bm{d} - u_{\mathrm{des}}(\bm{r}) \right|^2 \mathrm{d}\bm{r}.
\label{eq:obj_weight}
\end{align}
The function $\rho(\bm{r})$ is designed on the basis of the regional importance of the reproduction accuracy. However, in this study, we focus on the case of a uniform distribution, i.e., $\rho(\bm{r})=1$, for simplicity. 

\section{Sound Field Reproduction}

Several methods of approximately solving the minimization problem of Eq.~\eqref{eq:obj} have been proposed. We introduce two sound field reproduction methods, pressure matching and weighted mode matching, where weighted mode matching is a generalization of the mode matching method~\cite{Poletti:J_AES_2005,Daniel:AES114conv} previously proposed by the authors~\cite{Ueno:IEEE_ACM_J_ASLP2019}. 

\subsection{Pressure matching}

A simple strategy to solve the minimization problem of Eq.~\eqref{eq:obj} is to discretize the target region $\Omega$ into multiple control points, which is referred to as the pressure matching method. Assume that $N$ ($\ge L$) control points are placed on $\Omega$ and their positions are denoted by $\bm{r}_{\mathrm{c},n}$ ($n\in\{1,\ldots,N\}$). Equation~\eqref{eq:obj} is approximated as the minimization problem of the error between the synthesized and desired pressures at the control points. The optimization problem of pressure matching is described as
\begin{align}
 \minimize_{\bm{d}\in\mathbb{C}^L} \| \bm{Gd} - \bm{u}^{\mathrm{des}} \|^2,
\label{eq:obj_pm}
\end{align}
where $\bm{u}^{\mathrm{des}}=[u_{\mathrm{des}}(\bm{r}_{\mathrm{c},1}), \ldots, u_{\mathrm{des}}(\bm{r}_{\mathrm{c},N})]^{\mathsf{T}} \in \mathbb{C}^N$ is the vector of the desired sound pressures and $\bm{G}\in\mathbb{C}^{N \times L}$ is the transfer function matrix between $L$ secondary sources and $N$ control points given as
\begin{align}
 \bm{G} = 
\begin{bmatrix}
 g_1(\bm{r}_{\mathrm{c},1}) & \cdots & g_L(\bm{r}_{\mathrm{c},1}) \\
 \vdots & \ddots & \vdots \\
 g_1(\bm{r}_{\mathrm{c},N}) & \cdots & g_L(\bm{r}_{\mathrm{c},N}) 
\end{bmatrix}.
\end{align} 
Thus, the driving signal $\bm{d}$ for pressure matching is simply obtained by solving the least-squares problem Eq.~\eqref{eq:obj_pm} as
\begin{align}
 \bm{d} = \bm{G}^{\dagger} \bm{u}^{\mathrm{des}},
\end{align}
where $(\cdot)^{\dagger}$ represents the Moore--Penrose pseudoinverse. Since the computation of the inverse of $\bm{G}$ frequently becomes unstable, Eq.~\eqref{eq:obj_pm} is usually regularized as
\begin{align}
 \minimize_{\bm{d}\in\mathbb{C}^L} \| \bm{Gd} - \bm{u}^{\mathrm{des}} \|^2 + \eta \|\bm{d}\|^2,
\label{eq:obj_pm_reg}
\end{align}
where the second term is introduced to prevent an excessively large amplitude of $\bm{d}$ and $\eta$ is a constant regularization parameter. The solution of \eqref{eq:obj_pm_reg} is also simply obtained as
\begin{align}
 \bm{d} = \left( \bm{G}^{\mathsf{H}}\bm{G} + \eta \bm{I} \right)^{-1} \bm{G}^{\mathsf{H}} \bm{u}^{\mathrm{des}}.
\label{eq:pm}
\end{align}

Since the theoretical basis of the pressure matching method is the discretization of the regional integral in Eq.~\eqref{eq:obj}, the control points should basically be densely placed over $\Omega$. To reduce the number of control points (and also the number of secondary sources), several attempts have been made to optimize the placement of the secondary sources and control points~\cite{Koyama:IEEE_ACM_J_ASLP2020}.

\subsection{Weighted mode matching}

Weighted mode matching is a method of solving the minimization problem of Eq.~$\eqref{eq:obj}$ on the basis of the spherical wavefunction expansion of the sound field. A source-free interior sound field $u(\bm{r})$ can be expanded about the expansion center $\bm{r}_{\mathrm{o}}$ as~\cite{Martin:MultScat}
\begin{align}
 u(\bm{r}) = \sum_{\nu=0}^{\infty} \sum_{\mu=-\nu}^{\nu} \alpha_{\nu,\mu}(\bm{r}_{\mathrm{o}}) \varphi_{\nu,\mu}(\bm{r}-\bm{r}_{\mathrm{o}}),
\label{eq:sphexp}
\end{align}
where $\varphi_{\nu,\mu}(\bm{r})$ is the spherical wavefunction defined in the spherical coordinates $\bm{r}=(r,\theta,\phi)$ at the wave number $k=\omega/c$ with the sound speed $c$ as
\begin{align}
 \varphi_{\nu,\mu}(\bm{r}) = \sqrt{4\pi} j_{\nu}(kr) Y_{\nu,\mu}(\theta,\phi).
\label{eq:sphexp_basis}
\end{align}
Here, $j_{\nu}(\cdot)$ is the $\nu$th-order spherical Bessel function of the first kind, and $Y_{\nu,\mu}(\cdot)$ is the spherical harmonic function of order $\nu$ and degree $\mu$ defined as
\begin{align}
 Y_{\nu,\mu}(\theta,\phi) = \sqrt{\frac{2\nu +1}{4\pi}\frac{(\nu-\mu)!}{(\nu+\mu)!}} P_{\nu}^{\mu}(\cos\theta) \exp(\mathrm{j} \mu \phi), 
\end{align}
where $P_{\nu}^{\mu}(\cdot)$ is the associated Legendre function. Note that $\alpha_{0,0}(\bm{r})$ corresponds to $u(\bm{r})$.

First, the desired sound field $u_{\mathrm{des}}(\bm{r})$ and transfer function $g_l(\bm{r})$ are expanded about the expansion center $\bm{r}_{\mathrm{o}}$ as
\begin{align}
 u_{\mathrm{des}}(\bm{r}) &= \sum_{\nu=0}^{\infty} \sum_{\mu=-\nu}^{\nu} \tilde{u}_{\mathrm{des},\nu,\mu}(\bm{r}_{\mathrm{o}}) \varphi_{\nu,\mu}(\bm{r}-\bm{r}_{\mathrm{o}}) \\
 g_l(\bm{r}) &= \sum_{\nu=0}^{\infty} \sum_{\mu=-\nu}^{\nu} \tilde{g}_{l,\nu,\mu}(\bm{r}_{\mathrm{o}}) \varphi_{\nu,\mu}(\bm{r}-\bm{r}_{\mathrm{o}}).
\label{eq:sphexp_des_g}
\end{align}
By truncating the maximum order of the expansion in Eq.~\eqref{eq:sphexp_des_g} up to $N_{\mathrm{tr}}$,  $u_{\mathrm{des}}(\bm{r})$ and $\bm{g}(\bm{r})^{\mathsf{T}}$ in Eq.~\eqref{eq:obj} are approximated as
\begin{align}
 u_{\mathrm{des}}(\bm{r}) \approx \bar{\bm{\varphi}}(\bm{r}-\bm{r}_{\mathrm{o}})^{\mathsf{T}} \bm{b}(\bm{r}_{\mathrm{o}}) \\
 \bm{g}(\bm{r})^{\mathsf{T}} \approx \bar{\bm{\varphi}}(\bm{r}-\bm{r}_{\mathrm{o}})^{\mathsf{T}} \bm{C}(\bm{r}_{\mathrm{o}}), 
\end{align}
where $\bar{\bm{\varphi}}(\bm{r})\in\mathbb{C}^{(N_{\mathrm{tr}}+1)^2}$, $\bm{b}(\bm{r}_{\mathrm{o}})\in\mathbb{C}^{(N_{\mathrm{tr}}+1)^2}$, and $\bm{C}(\bm{r}_{\mathrm{o}})\in\mathbb{C}^{(N_{\mathrm{tr}}+1)^2 \times L}$ are the vectors and matrix consisting of $\varphi_{\nu,\mu}(\bm{r}-\bm{r}_{\mathrm{o}})$, $\tilde{u}_{\mathrm{des}}(\bm{r}_{\mathrm{o}})$, and $\tilde{g}_l(\bm{r}_{\mathrm{o}})$, respectively. Thus, the objective function $J$ is approximated as
\begin{align}
 J &\approx \int_{\Omega} \left| \bar{\bm{\varphi}}(\bm{r})^{\mathsf{T}} \left( \bm{C}\bm{d} - \bm{b} \right) \right|^2 \mathrm{d}\bm{r} \notag\\
&= \left( \bm{C}\bm{d} - \bm{b} \right)^{\mathsf{H}} \int_{\Omega} \bar{\bm{\varphi}}(\bm{r})^{\ast} \bar{\bm{\varphi}}(\bm{r})^{\mathsf{T}} \mathrm{d}\bm{r} \left( \bm{C}\bm{d} - \bm{b} \right) \notag\\
&= \left( \bm{C}\bm{d} - \bm{b} \right)^{\mathsf{H}} \bm{W} \left( \bm{C}\bm{d} - \bm{b} \right), 
\label{eq:obj_app_wmm}
\end{align}
where the $(i,j)$th element of $\bm{W}\in\mathbb{C}^{(N_{\mathrm{tr}}+1)^2 \times (N_{\mathrm{tr}}+1)^2}$ is defined as
\begin{align}
 (\bm{W})_{i,j} = \int_{\Omega} \varphi_{\nu_i,\mu_i}(\bm{r})^{\ast} \varphi_{\nu_j,\mu_j}(\bm{r}) \mathrm{d}\bm{r}.
\label{eq:weight_wmm}
\end{align}
Here, the expansion center $\bm{r}_{\mathrm{o}}$ is omitted for notational simplicity. The matrix $\bm{W}$ is the weighting factor for the expansion coefficients of the spherical wavefunctions. Each element of $\bm{W}$ must be computed by numerical integration for an arbitrary shape of $\Omega$. There are some techniques to avoid the numerical integration when $\Omega$ has a simple shape such as a sphere~\cite{Ueno:IEEE_ACM_J_ASLP2019}. When the expected regional accuracy $\rho(\bm{r})$ in Eq.~\eqref{eq:obj_weight} is used, the element of the weighting matrix becomes
\begin{align}
 (\bm{W})_{i,j} = \int_{\Omega} \rho(\bm{r}) \varphi_{\nu_i,\mu_i}(\bm{r})^{\ast} \varphi_{\nu_j,\mu_j}(\bm{r}) \mathrm{d}\bm{r}.
\label{eq:weight_wmm_rho}
\end{align}

As in pressure matching, the optimization problem for weighted mode matching is formulated by adding the regularization term to Eq.~\eqref{eq:obj_app_wmm} as
\begin{align}
 \minimize_{\bm{d}\in\mathbb{C}^L} \left( \bm{C}\bm{d} - \bm{b} \right)^{\mathsf{H}} \bm{W} \left( \bm{C}\bm{d} - \bm{b} \right) + \lambda \|\bm{d}\|^2,
\end{align}
where $\lambda$ is a constant parameter. This problem also has the closed-form solution 
\begin{align}
 \bm{d} = \left( \bm{C}^{\mathsf{H}} \bm{W} \bm{C} + \lambda \bm{I} \right)^{-1} \bm{C}^{\mathsf{H}} \bm{W} \bm{b}.
\label{eq:wmm}
\end{align}
When the weighting matrix $\bm{W}$ is the identity matrix, Eq.~\eqref{eq:wmm} corresponds to the driving signal of the mode matching method, which requires an appropriate setting of the truncation order $N_{\mathrm{tr}}$ for the spherical wavefunction expansion because it is sensitive to the reproduction performance and numerical stability. On the other hand, in weighted mode matching, the weighting factor for each expansion coefficient is determined by regional integration in Eq.~\eqref{eq:weight_wmm}. Therefore, the maximum truncation order can usually be set at a sufficiently large number. 

In pressure matching, the parameters to be made to coincide are the sound pressures at the control points. They are converted into the expansion coefficients in weighted mode matching to approximate Eq.~\eqref{eq:obj}. This transformation enables us to reduce the number of parameters to be made to coincide as well as the number of basis functions as shown in Ref.~\cite{Ueno:IEEE_ACM_J_ASLP2019}. However, it is necessary to obtain the expansion coefficients of the desired sound field $\bm{b}$ and transfer functions $\bm{C}$. If they can be simply modeled by an analytical function, such as a plane or spherical wave, $\bm{b}$ and $\bm{C}$ can be computed on the basis of the positions of $\Omega$ and the secondary sources. In many practical situations, $\bm{b}$ and/or $\bm{C}$ must be estimated by using microphone measurements. 

\section{Estimation of Expansion Coefficients of Spherical Wavefunctions}

To reproduce a captured sound field by using the weighted mode matching method, the expansion coefficients of the desired sound field must be estimated in the recording area. In addition, when the transfer functions of the secondary sources cannot be simply modeled, e.g., owing to the reverberation, their expansion coefficients must be obtained in advance. We here introduce a method of estimating the expansion coefficients of spherical wavefunctions from microphone measurements that is applicable to obtain both $\bm{b}$ and $\bm{C}$. 

Most of the methods of estimating the expansion coefficients of spherical wavefunctions require that the microphones are placed on a spherical surface. Moreover, the maximum expansion order must be set for truncation in an empirical manner. The restriction on the microphone array geometry significantly reduces the practical applicability of these methods. For example, when a large region of the sound field must be captured, it is sometimes infeasible to implement and place a large spherical microphone array. We previously proposed a method based on the infinite-dimensional harmonic analysis, which allows to use arbitrarily placed microphones. Moreover, no empirical truncation of expansion order is necessary. 

\subsection{Translation operator and its properties}

First, we introduce the translation operator for the expansion coefficients of the spherical wavefunctions. Equation~\eqref{eq:sphexp} can be represented in a matrix form as 
\begin{align}
 u(\bm{r}) = \bm{\varphi}(\bm{r}-\bm{r}_{\mathrm{o}})^{\mathsf{T}} \bm{\alpha}(\bm{r}_{\mathrm{o}}), 
\end{align}
where $\bm{\varphi}(\bm{r}-\bm{r}_{\mathrm{o}})\in\mathbb{C}^{\infty}$ and $\bm{\alpha}(\bm{r}_{\mathrm{o}})\in\mathbb{C}^{\infty}$ are infinite-dimensional vectors of the spherical wavefunctions $\varphi_{\nu,\mu}(\bm{r}-\bm{r}_{\mathrm{o}})$ and expansion coefficients $\alpha_{\nu,\mu}(\bm{r}_{\mathrm{o}})$, respectively. The translation operator $\bm{T}(\bm{r}-\bm{r}^{\prime})\in\mathbb{C}^{\infty\times\infty}$ relates the expansion coefficients about two different expansion centers $\bm{r}$ and $\bm{r}^{\prime}$, $\bm{\alpha}(\bm{r})$ and $\bm{\alpha}(\bm{r}^{\prime})$, as~\cite{Martin:MultScat}
%The expansion coefficients around two expansion centers $\bm{r}$ and $\bm{r}^{\prime}$, $\bm{\alpha}(\bm{r})$ and $\bm{\alpha}(\bm{r}^{\prime})$, can be related as~\cite{Martin:MultScat}
\begin{align}
 \bm{\alpha}(\bm{r}) = \bm{T}(\bm{r}-\bm{r}^{\prime}) \bm{\alpha}(\bm{r}^{\prime}),
\end{align}
where the element corresponding to order $\nu^{\prime}$ and degree $\mu^{\prime}$ of $\bm{T}(\bm{r})\bm{\alpha}$, denoted as $[\bm{T}(\bm{r})\bm{\alpha}]_{\nu^{\prime},\mu^{\prime}}$, is defined as
\begin{align}
 &[\bm{T}(\bm{r})\bm{\alpha}]_{\nu^{\prime},\mu^{\prime}} = \sum_{\nu=0}^{\infty} \sum_{\mu=-\nu}^{\nu} \left[ 4\pi (-1)^{\mu} \mathrm{j}^{\nu^{\prime}-\nu} \right. \notag\\
 & \ \ \left. \cdot\sum_{l=0}^{\nu^{\prime}+\nu} \mathrm{j}^l j_l(kr) Y_{l,\mu-\mu^{\prime}}(\theta,\phi)^{\ast} \mathcal{G}(\nu, \mu; \nu^{\prime}, -\mu^{\prime}; l) \right] \alpha_{\nu,\mu}.
\end{align}
Here, $\mathcal{G}(\cdot)$ is the Gaunt coefficient. 
%Here, $[\cdot]_{\nu^{\prime},\mu^{\prime}}$ is the element corresponding to order $\nu^{\prime}$ and degree $\mu^{\prime}$, and $\mathcal{G}(\cdot)$ is the Gaunt coefficient. 
The translation operator $\bm{T}(\bm{r}-\bm{r}^{\prime})$ has the following important properties:
%The translation operator satisfies the following equations:
\begin{align}
 \bm{T}(-\bm{r}) &= \bm{T}^{-1}(\bm{r}) = \bm{T}(\bm{r})^{\mathsf{H}} \label{eq:trans_inv}\\
 \bm{T}(\bm{r}+\bm{r}^{\prime}) &= \bm{T}(\bm{r})\bm{T}(\bm{r}^{\prime}). \label{eq:trans_shift}
\end{align}

\subsection{Infinite-dimensional harmonic analysis}

Suppose that the source-free target capturing region $\Omega$, which has the same shape as the target reproducing region $\Omega$ in Fig.~\ref{fig:sfc}, is set. We consider how to estimate the expansion coefficients of spherical wavefunctions of the sound field at the position $\bm{r}\in\Omega$ by using microphones arbitrarily placed in $\Omega$. The microphone directivity patterns are assumed to be given as their expansion coefficients $c_{m,\nu,\mu}$, where $m$ ($\in\{1.\ldots,M\}$) denotes the microphone index.  
%The expansion coefficients $c_{m,\nu,\mu}$ denote of the directivity pattern of the $m$th microphone ($m\in\{1.\ldots,M\}$) are denoted by $c_{m,\nu,\mu}$. 
Then, the observation of the $m$th microphone at $\bm{r}_{\mathrm{m},m}$, denoted by $s_m$, is described as the inner product of $c_{m,\nu,\mu}$ and $\alpha_{\nu,\mu}(\bm{r}_{\mathrm{m},m})$:
\begin{align}
 s_m &= \sum_{\nu=0}^{\infty} \sum_{\mu=-\nu}^{\nu} c_{m,\nu,\mu} \alpha_{\nu,\mu}(\bm{r}_{\mathrm{m},m}) \notag\\
&= \bm{c}_{m}^{\mathsf{H}} \bm{\alpha}(\bm{r}_{\mathrm{m},m}) \notag\\
&= \bm{c}_{m}^{\mathsf{H}} \bm{T}(\bm{r}_{\mathrm{m},m}-\bm{r}) \bm{\alpha}(\bm{r}),
\label{eq:s_i}
\end{align}
where $\bm{c}_m\in\mathbb{C}^{\infty}$ is the infinite-dimensional vector of $c_{m,\nu,\mu}$. Equation~\eqref{eq:s_i} can be rewritten as
\begin{align}
 \bm{s} = \bm{\Xi}(\bm{r})^{\mathsf{H}} \bm{\alpha}(\bm{r}),
\end{align}
where $\bm{\Xi}(\bm{r})\in\mathbb{C}^{\infty \times M}$ is obtained as
\begin{align}
 \bm{\Xi}(\bm{r}) &= \left[ (\bm{c}_1^{\mathsf{H}}\bm{T}(\bm{r}_{\mathrm{m},1}-\bm{r}))^{\mathsf{H}}, \cdots, (\bm{c}_M^{\mathsf{H}}\bm{T}(\bm{r}_{\mathrm{m},M}-\bm{r}))^{\mathsf{H}} \right] \notag\\
&= \left[ \bm{T}(\bm{r}_{\mathrm{m},1}-\bm{r})\bm{c}_1, \cdots, \bm{T}(\bm{r}_{\mathrm{m},M}-\bm{r})\bm{c}_M \right] .
\end{align}
Here, the property of the translation operator \eqref{eq:trans_inv} is used. The expansion coefficients $\bm{\alpha}(\bm{r})$ are estimated as
\begin{align}
 \bm{\alpha}(\bm{r}) = \bm{\Xi}(\bm{r}) (\bm{\Psi} + \xi \bm{I})^{-1}\bm{s},
\label{eq:est_coeff}
\end{align}
where $\xi$ is a constant parameter and $\bm{\Psi}=\bm{\Xi}(\bm{r})^{\mathsf{H}}\bm{\Xi}(\bm{r})\in\mathbb{C}^{M \times M}$. From the property in Eq.~\eqref{eq:trans_shift}, the $(m,m^{\prime})$th element of $\bm{\Psi}$ becomes
\begin{align}
 (\bm{\Psi})_{m,m^{\prime}} &= \bm{c}_m^{\mathsf{H}} \bm{T}(\bm{r}_{\mathrm{m},m}-\bm{r}) \bm{T}(\bm{r}-\bm{r}_{\mathrm{m},m^{\prime}}) \bm{c}_{m^{\prime}} \notag\\
&= \bm{c}_m^{\mathsf{H}} \bm{T}(\bm{r}_{\mathrm{m},m}-\bm{r}_{\mathrm{m},m^{\prime}}) \bm{c}_{m^{\prime}}.
\label{eq:psi}
\end{align}
Therefore, $\bm{\Psi}$ does not depend on the position $\bm{r}$ and depends only on the microphone positions and directivities. Since the microphone directivity $\bm{c}_m$ is typically modeled by low-order coefficients, \eqref{eq:psi} can be simply computed in practice. 

In the sound field recording, it is frequently impractical to capture the sound field in a large region by using a single large spherical microphone array. Our proposed estimation method allows to use arbitrarily placed microphones, for example, distributed small microphone arrays. Such a sound field recording system will be useful in practical situations because of its flexibility and scalability. 

\section{Experiments}

\begin{figure}[t]
 \centerline{\includegraphics[width=0.65\columnwidth]{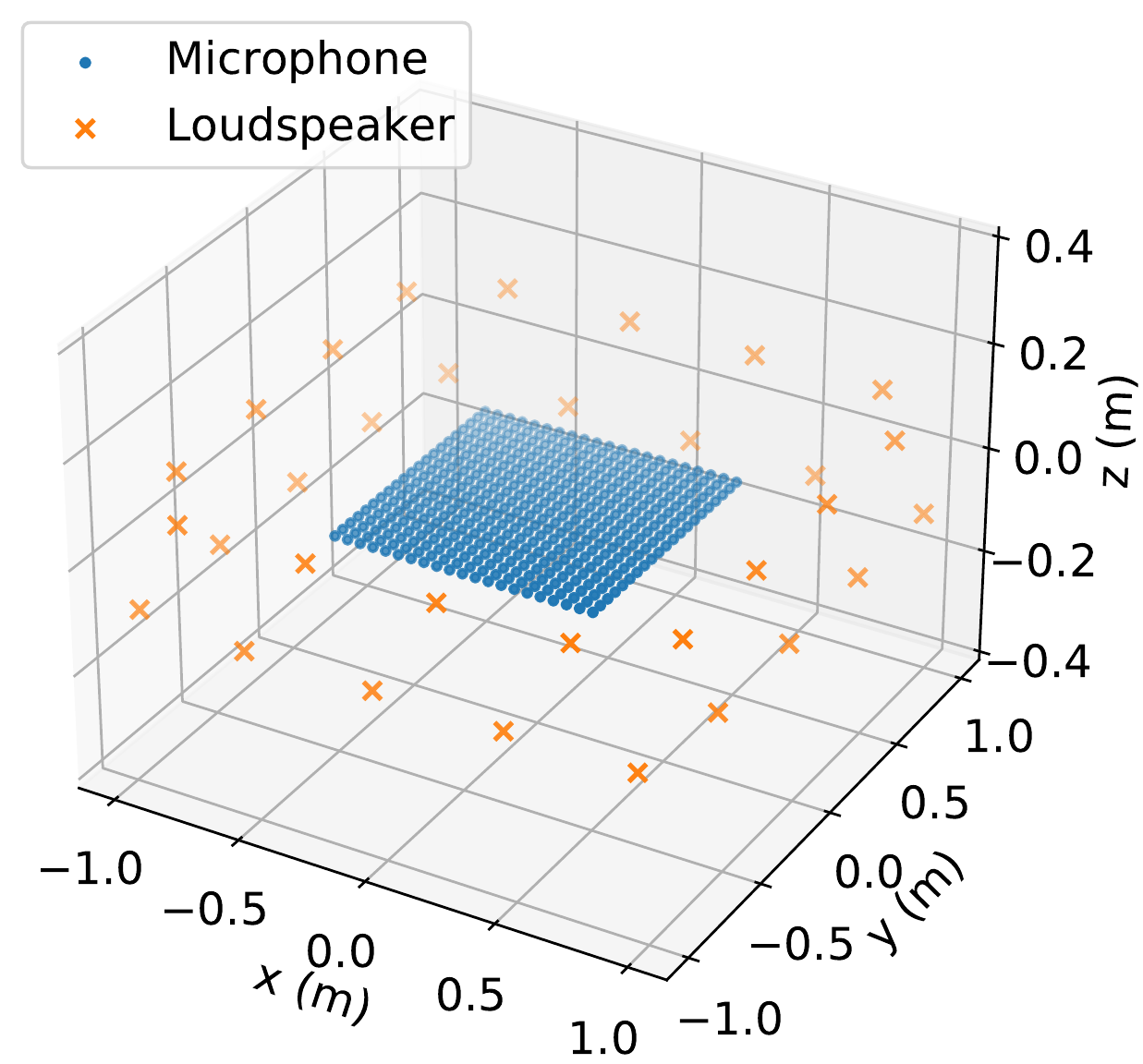}}
 \caption{Positions of loudspeakers and evaluation points.}
\label{fig:exp_pos}
\end{figure}
\begin{figure}[t]
 \centerline{\includegraphics[width=0.75\columnwidth]{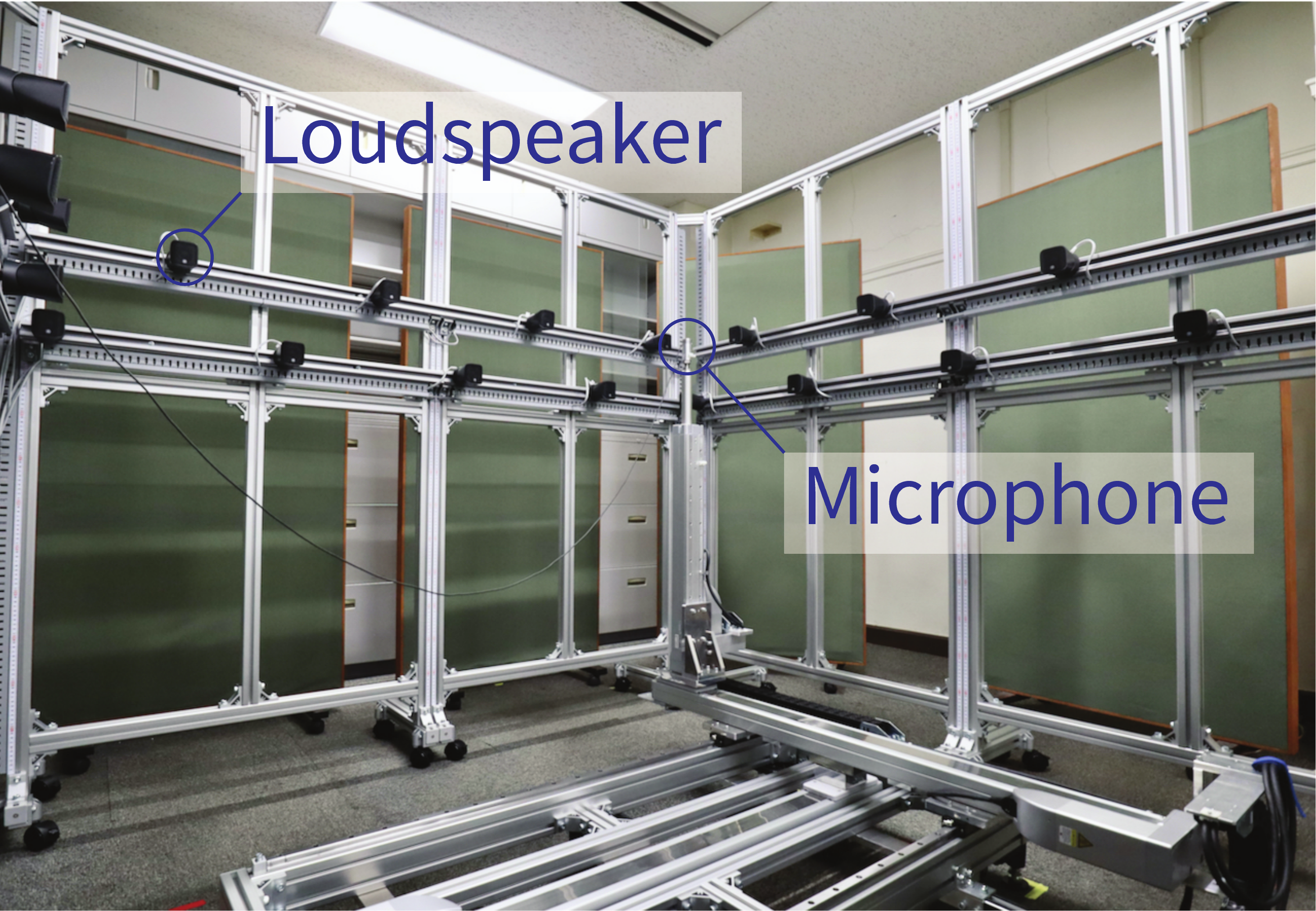}}
 \caption{Impulse response measurement system.}
\label{fig:exp_meas}
\end{figure}

%\begin{figure*}[t]
% \centerline{
% \subfloat[True]{\includegraphics[width=0.7\columnwidth]{fig/exp/control_true.pdf}}
% \subfloat[Pressure matching]{\includegraphics[width=0.7\columnwidth]{fig/exp/control_syn_pm.pdf}}
% \subfloat[Weighted mode matching]{\includegraphics[width=0.7\columnwidth]{fig/exp/control_syn_wmm.pdf}}
% }
% \caption{True and reproduced pressure distributions at $t=0.51~\mathrm{s}$. Black crosses indicate microphone positions.}
%\label{fig:exp_pres}
%%\end{figure*}
%%\begin{figure*}[t]
% \centerline{
% \hspace{0.7\columnwidth}
% \subfloat[Pressure matching]{\includegraphics[width=0.7\columnwidth]{fig/exp/control_error_pm.pdf}}
% \subfloat[Weighted mode matching]{\includegraphics[width=0.7\columnwidth]{fig/exp/control_error_wmm.pdf}}
% }
% \caption{Time-averaged square error distributions. SDRs were 3.85~dB for pressure matching and 4.65~dB for weighted mode matching.}
%\label{fig:exp_error}
%\end{figure*}

\begin{figure}[t]
 \centerline{
 \subfloat[True]{\includegraphics[width=0.55\columnwidth,clip]{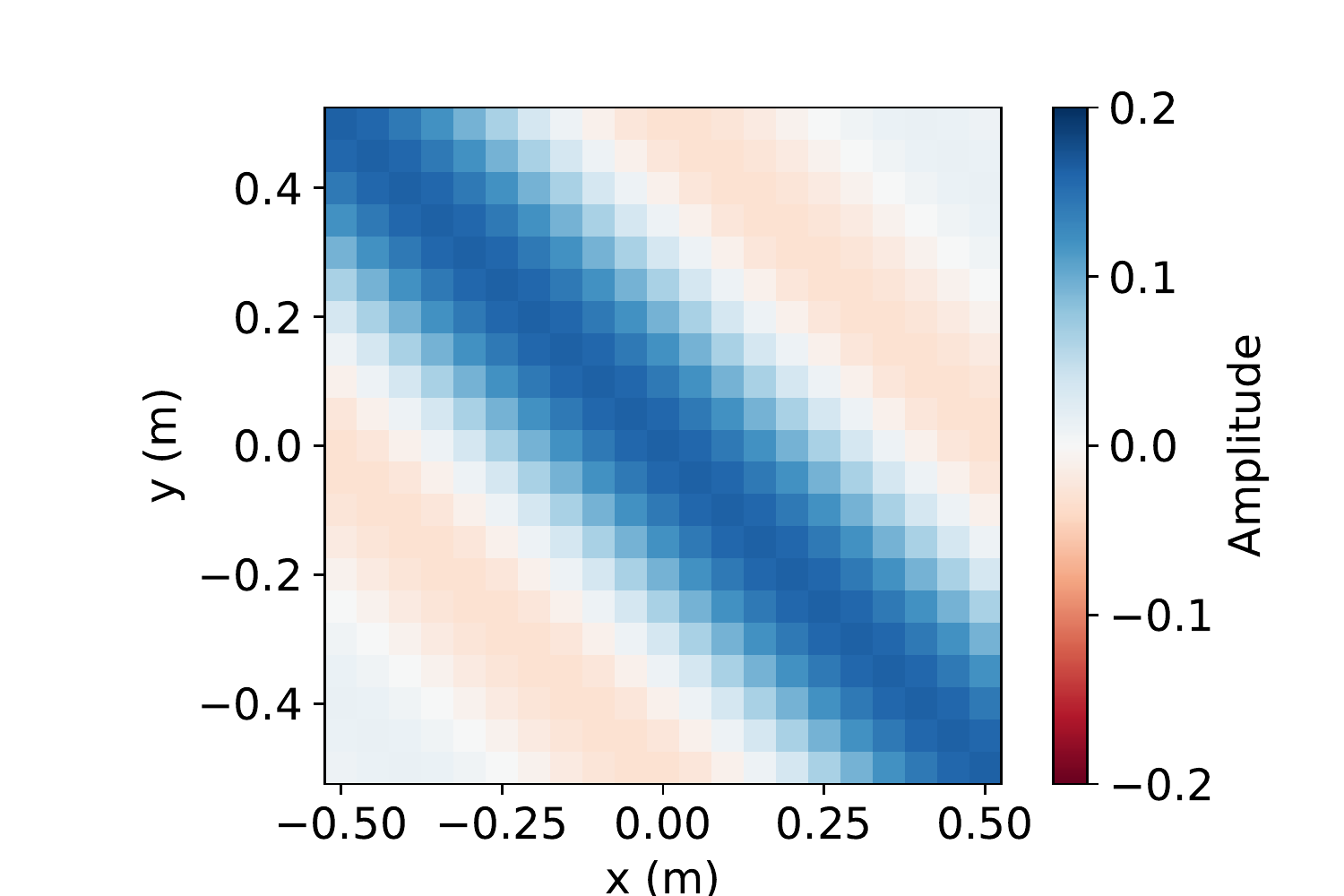}}}
 \centerline{
 \subfloat[Pressure matching]{\includegraphics[width=0.55\columnwidth,clip]{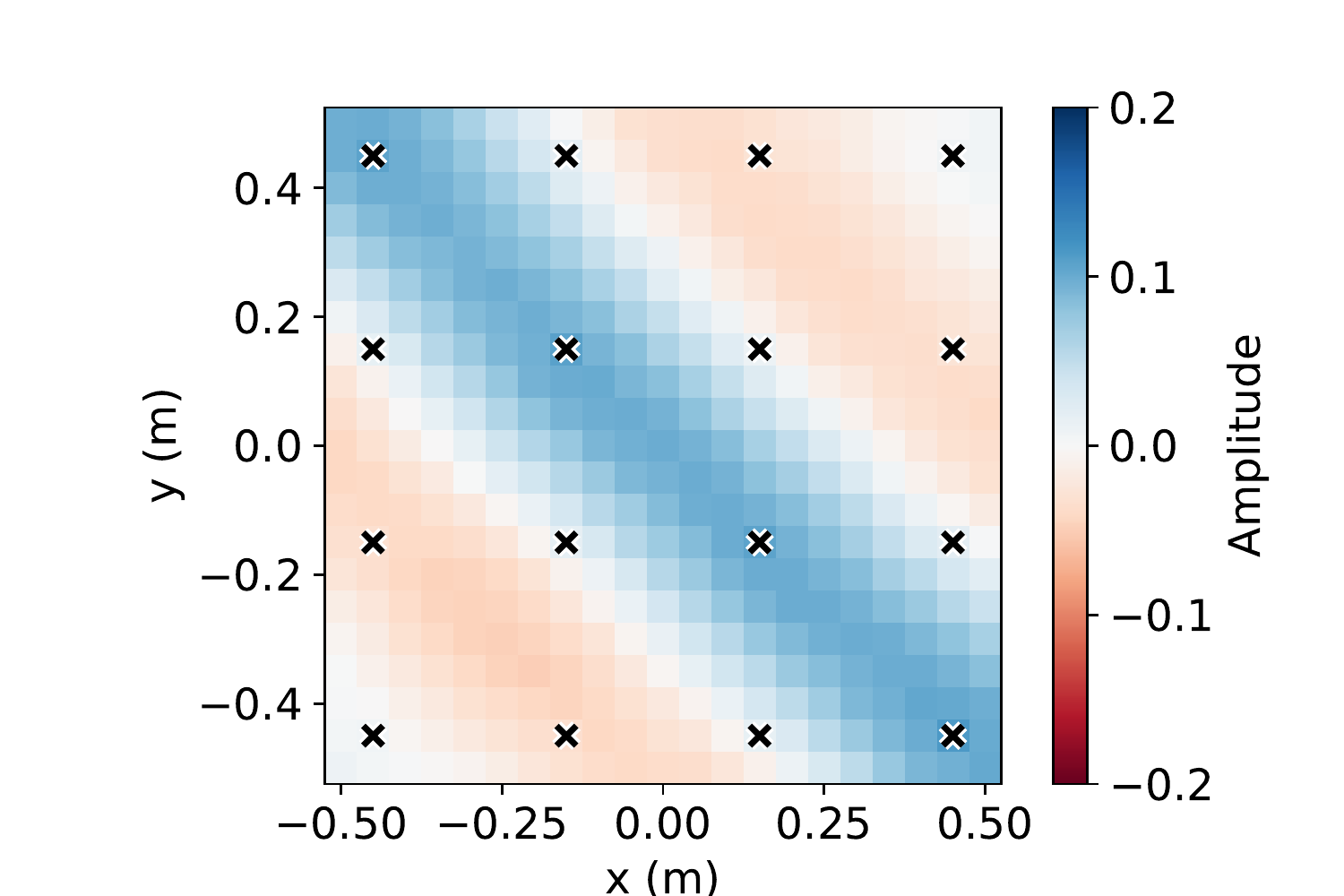}}
 \hspace{-20pt}
 \subfloat[Weighted mode matching]{\includegraphics[width=0.55\columnwidth,clip]{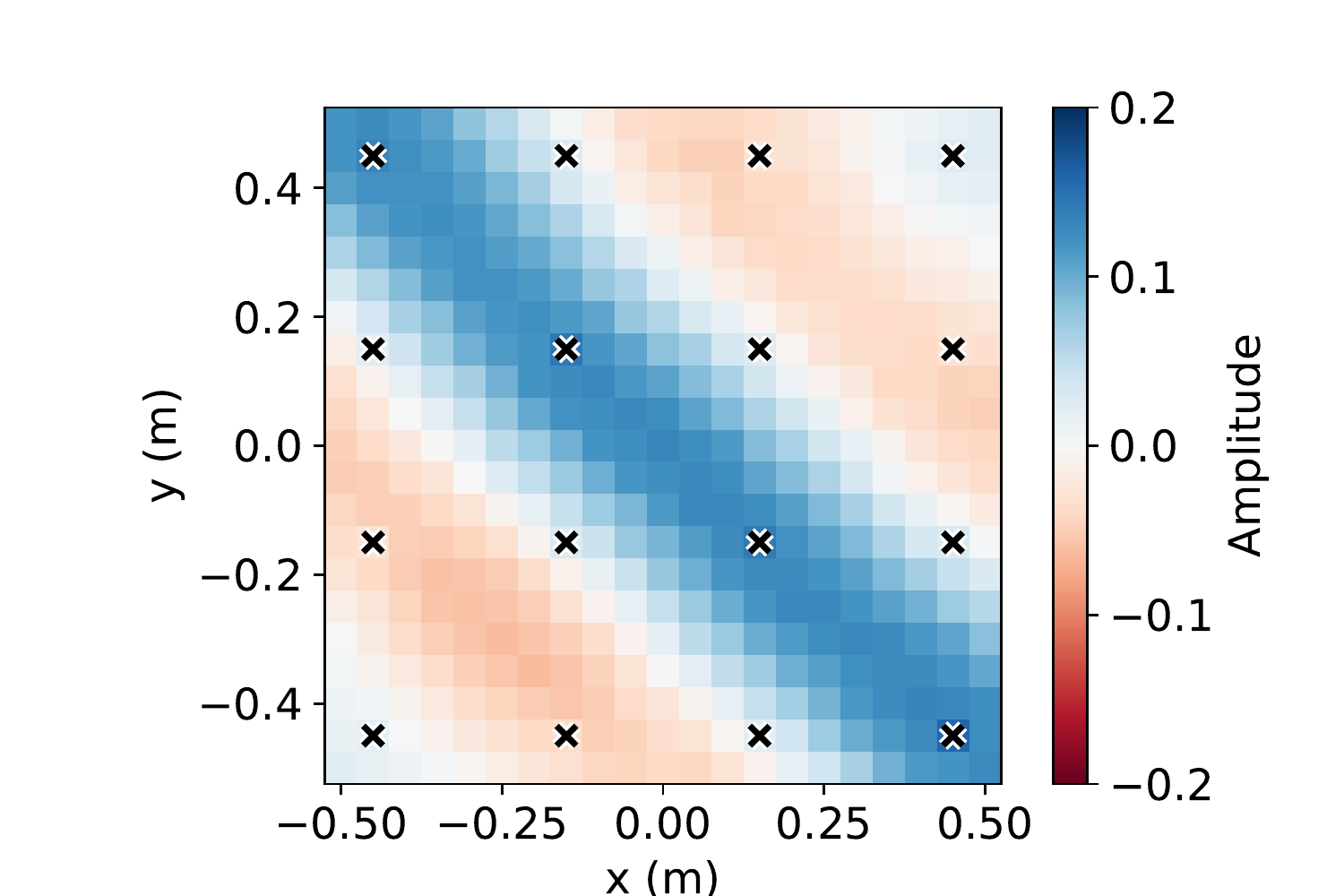}}
 }
 \caption{True and reproduced pressure distributions at $t=0.51~\mathrm{s}$. Black crosses indicate microphone positions.}
\label{fig:exp_pres}
%\end{figure}
%\begin{figure}[t]
 \centerline{
 \subfloat[Pressure matching]{\includegraphics[width=0.55\columnwidth,clip]{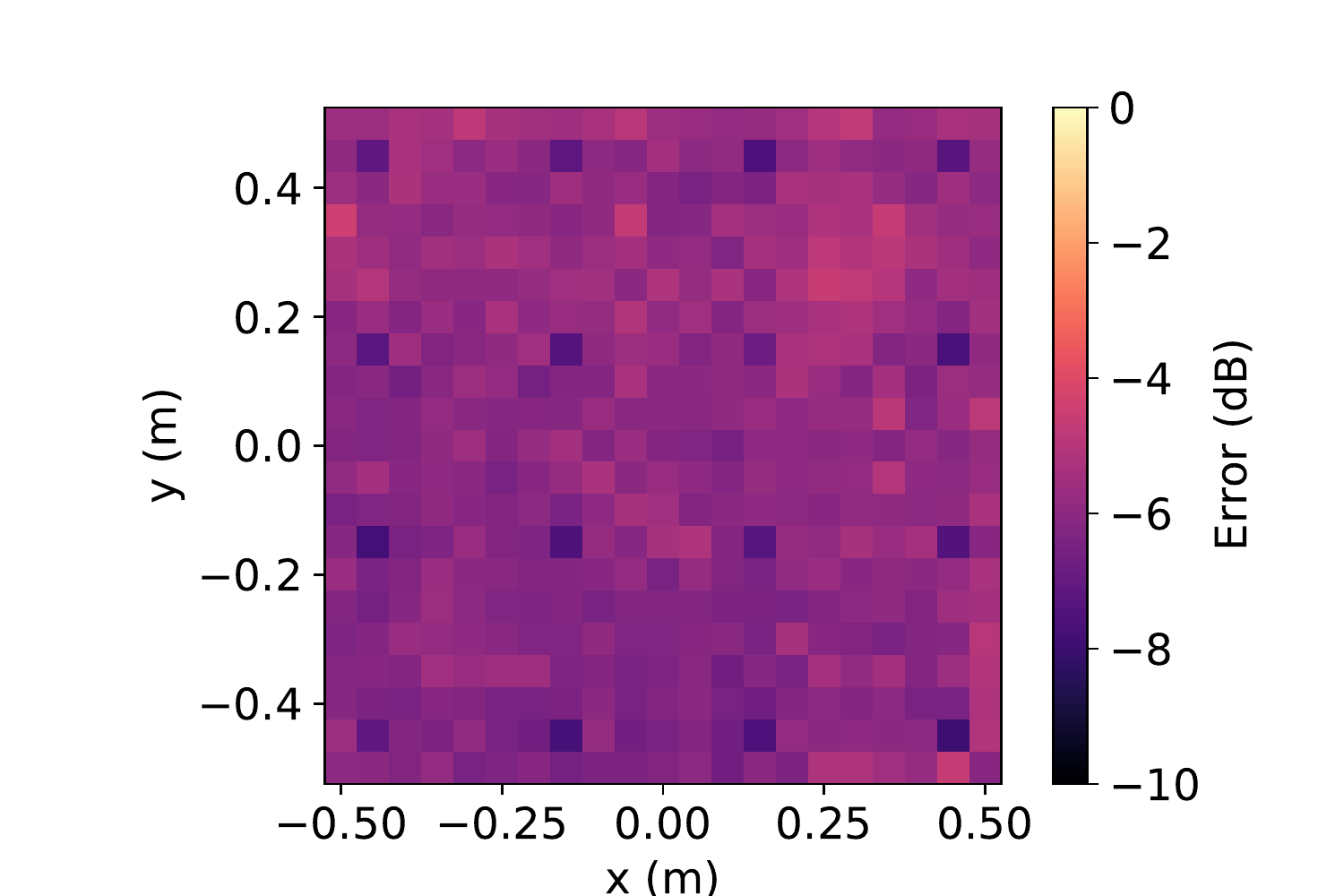}}
\hspace{-20pt}
 \subfloat[Weighted mode matching]{\includegraphics[width=0.55\columnwidth,clip]{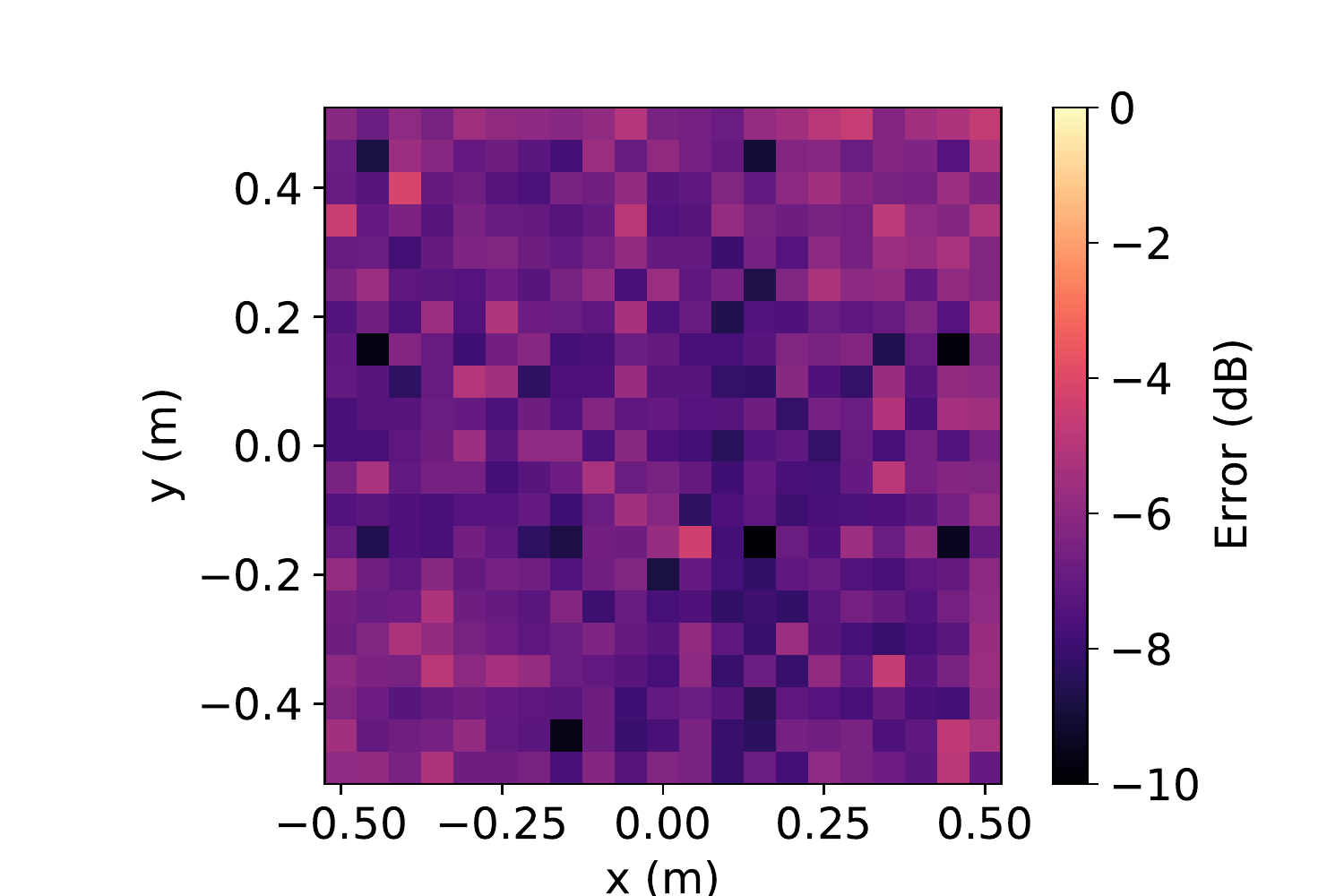}}
 }
 \caption{Time-averaged square error distributions. SDRs were 3.85~dB for pressure matching and 4.65~dB for weighted mode matching.}
\label{fig:exp_error}
\end{figure}

\begin{figure}[t]
 \centerline{\includegraphics[width=0.9\columnwidth]{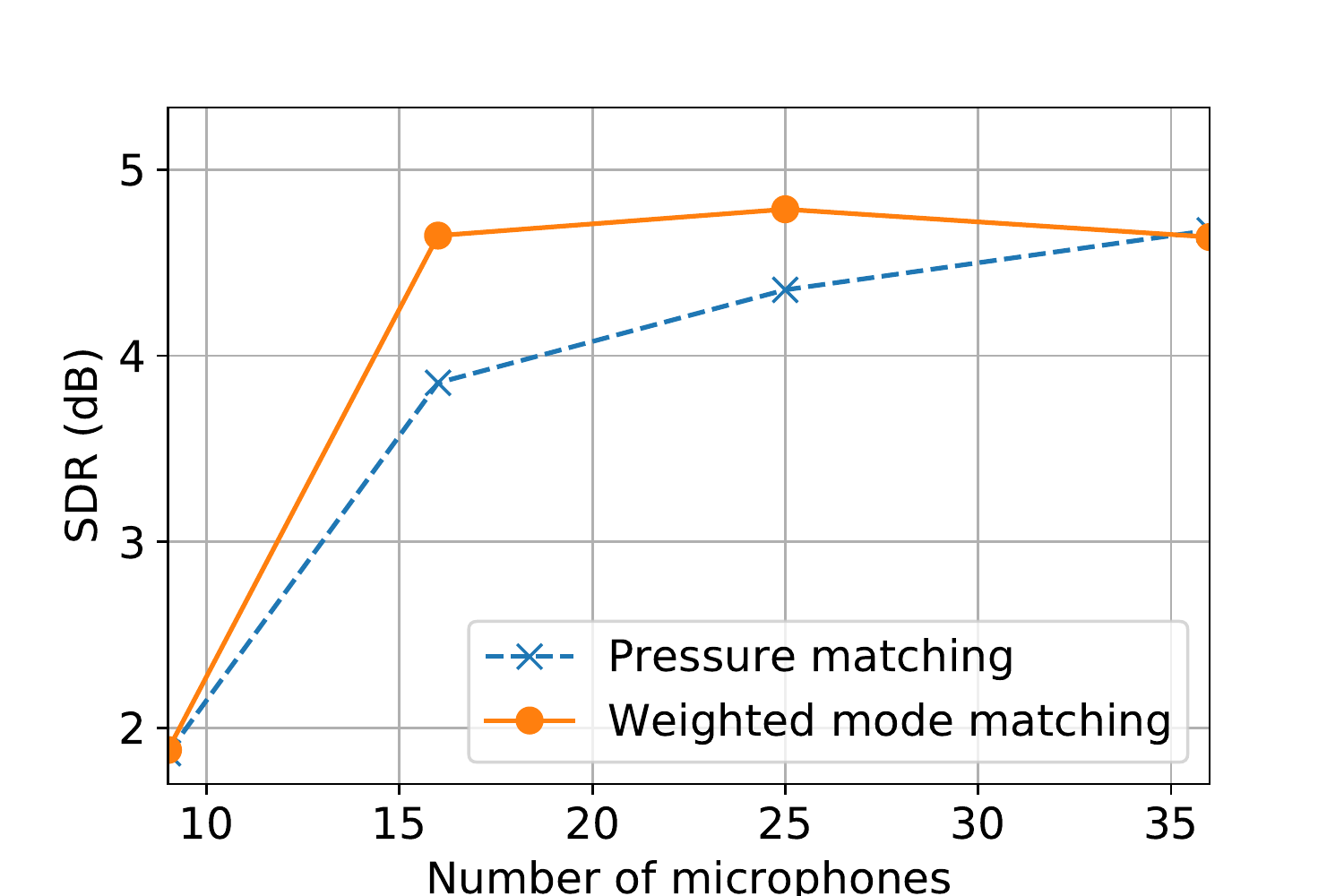}}
 \caption{SDR with respect to number of microphones.}
\label{fig:exp_sdr}
\end{figure}

We conducted experiments using impulse responses measured in a practical environment, where the recently published impulse response dataset MeshRIR was used~\cite{Koyama:WASPAA2021}. The positions of the loudspeakers and evaluation points are shown in in Fig.~\ref{fig:exp_pos}. Along the borders of two squares with dimensions of $2.0~\mathrm{m} \times 2.0~\mathrm{m}$ at heights of $z=-0.2~\mathrm{m}$ and $0.2~\mathrm{m}$, 32 loudspeakers were regularly placed; therefore, 16 loudspeakers were placed along each square. We used ordinary closed loudspeakers (YAMAHA, VXS1MLB).
%32 loudspeakers were regularly placed along the borders of two squares with dimensions of $2.0~\mathrm{m} \times 2.0~\mathrm{m}$ at heights of $z=-0.2~\mathrm{m}$ and $0.2~\mathrm{m}$. Small closed loudspeakers (YAMAHA, VXS1MLB) were used. 
The measurement region was a square with dimensions of $1.0~\mathrm{m} \times 1.0~\mathrm{m}$ at $z=0.0~\mathrm{m}$. The measurement region was discretized at intervals of $0.05~\mathrm{m}$, and $21 \times 21$ ($=441$) evaluation points were obtained. We measured impulse response at each evaluation point with an omnidirectional microphone (Primo, EM272J) equipped on a Cartesian robot (see Fig.~\ref{fig:exp_meas}). The excitation signal of impulse response measurement was linear swept-sine signal~\cite{Suzuki:JASA1995}. The details of the measurement conditions are described in~\cite{Koyama:WASPAA2021}.
%At each evaluation point, we measured impulse response by using a Cartesian robot with an omnidirectional microphone (Primo, EM272J) to control the measurement position in three dimensions (see Fig.~\ref{fig:exp_meas}). Linear swept-sine signals were used for the excitation signal of impulse response measurement~\cite{Suzuki:JASA1995}. 
The sampling frequency of the impulse responses was $48~\mathrm{kHz}$, but it was downsampled to $8~\mathrm{kHz}$.

We compared pressure matching and weighted mode matching. The target region was the same as the region of the evaluation points. The microphone positions were chosen from the evaluation points, which were used as control points in pressure matching and to estimate the expansion coefficients of the transfer functions in weighted mode matching. In weighted mode matching, the expansion coefficients were estimated up to the 12th order, and the regularization parameter in Eq.~\eqref{eq:est_coeff} was set as $10^{-3}$. 
We set the desired sound field to a single plane wave arriving from $(\theta,\phi)=(\pi/2,\pi/4)$. 
%The desired sound field was a single plane wave whose arrival direction was $(\theta,\phi)=(\pi/2,\pi/4)$. 
The source signal was a pulse signal whose frequency band was low-pass-filtered up to $700~\mathrm{Hz}$.
%The source signal was a low-pass-filtered pulse signal, where the cutoff frequency was $700~\mathrm{Hz}$. 
The regularization parameters $\eta$ in Eq.~\eqref{eq:pm} and $\lambda$ in Eq.~\eqref{eq:wmm} were set as $10^2$ and $10^0$, respectively, which were the best values chosen from the range $[10^{-2}, 10^{3}]$ regularly discretized in a logarithmic scale. The filter for obtaining the driving signals was designed in the time domain, and its length was $8192~\mathrm{samples}$.

For the evaluation measure, we define the signal-to-distortion ratio (SDR) as 
\begin{align}
 \mathrm{SDR} = \frac{\iint |u_{\mathrm{des}}(\bm{r},t)|^2 \mathrm{d}\bm{r}\mathrm{d}t}{\iint |u_{\mathrm{syn}}(\bm{r},t) - u_{\mathrm{des}}(\bm{r},t)|^2 \mathrm{d}\bm{r}\mathrm{d}t},
\end{align}
where the integration was computed at the evaluation points, except the control points, for $\bm{r}$ with a signal length of $8192~\mathrm{samples}$ for $t$. 

Figure~\ref{fig:exp_pres} shows the true and reproduced pressure distributions for pressure matching and weighted mode matching at $t=0.51~\mathrm{s}$. Black crosses indicate microphone positions. Sixteen ($=4 \times 4$) microphones were regularly placed over the target region. Time-averaged square error distributions are shown in Fig.~\ref{fig:exp_error}. In pressure matching, a small time-averaged square error was achieved at the microphone positions, i.e., control points. On the other hand, that in weighted mode matching was small in the entire target region. SDRs were $3.85~\mathrm{dB}$ for the pressure matching and $4.65~\mathrm{dB}$ for weighted mode matching. 

The SDRs obtained using $9$ ($=3 \times 3$), $16$ ($=4 \times 4$), $25$ ($=5 \times 5$), and $36$ ($=6 \times 6$) microphones are plotted in Fig.~\ref{fig:exp_sdr}. Although the reproduction accuracies of the two methods were almost the same in the case of 36 microphones, the difference increased in the cases of 16 and 25 microphones. Therefore, weighted mode matching performs better when the number of available microphones is small. A similar tendency was also observed in simulation experiments~\cite{Ueno:IEEE_ACM_J_ASLP2019}.

\section{Conclusion}

A sound field reproduction method, weighted mode matching, was experimentally evaluated in a practical environment. Weighted mode matching is a generalization of the mode matching method, which is a method based on the spherical wavefunction expansion of a sound field. The weighting factor of expansion coefficients is obtained on the basis of the setting of the target region and expected regional accuracy. To estimate the expansion coefficients of the transfer functions of the loudspeakers and the desired sound field, we introduced the infinite-dimensional harmonic analysis method. Experiments were conducted to compare the performance of weighted mode matching with that of pressure matching using MeshRIR. Weighted mode matching achieved high reproduction accuracy when the number of microphones was small. 

\bibliographystyle{IEEEtran}
\bibliography{str_def_abrv,koyama_en,refs}

% Generated by IEEEtran.bst, version: 1.12 (2007/01/11)
\begin{thebibliography}{10}
\providecommand{\url}[1]{#1}
\csname url@samestyle\endcsname
\providecommand{\newblock}{\relax}
\providecommand{\bibinfo}[2]{#2}
\providecommand{\BIBentrySTDinterwordspacing}{\spaceskip=0pt\relax}
\providecommand{\BIBentryALTinterwordstretchfactor}{4}
\providecommand{\BIBentryALTinterwordspacing}{\spaceskip=\fontdimen2\font plus
\BIBentryALTinterwordstretchfactor\fontdimen3\font minus
  \fontdimen4\font\relax}
\providecommand{\BIBforeignlanguage}[2]{{%
\expandafter\ifx\csname l@#1\endcsname\relax
\typeout{** WARNING: IEEEtran.bst: No hyphenation pattern has been}%
\typeout{** loaded for the language `#1'. Using the pattern for}%
\typeout{** the default language instead.}%
\else
\language=\csname l@#1\endcsname
\fi
#2}}
\providecommand{\BIBdecl}{\relax}
\BIBdecl

\bibitem{Berkhout:JASA1980}
A.~J. Berkhout, D.~{de Vries}, and M.~M. Boone, ``A new method to acquire
  impulse responses in concert halls,'' \emph{J. Acoust. Soc. Amer.}, vol.~68,
  no.~1, pp. 179--183, 1998.

\bibitem{Spors:AES124conv}
S.~Spors, R.~Rabenstein, and J.~Ahrens, ``The theory of wave field synthesis
  revisited,'' in \emph{Proc. 124th {AES} Conv.}, Amsterdam, Netherlands, 2008.

\bibitem{Poletti:J_AES_2005}
M.~A. Poletti, ``Three-dimensional surround sound systems based on spherical
  harmonics,'' \emph{J. Audio Eng. Soc.}, vol.~53, no.~11, pp. 1004--1025,
  2005.

\bibitem{Ahrens:Acustica2008}
J.~Ahrens and S.~Spors, ``An analytical approach to sound field reproduction
  using circular and spherical loudspeaker distributions,'' \emph{Acta Acustica
  united with Acustica}, vol.~94, pp. 988--999, 2008.

\bibitem{Wu:IEEE_J_ASLP2009}
Y.~J. Wu and T.~D. Abhayapala, ``Theory and design of soundfield reproduction
  using continuous loudspeaker concept,'' \emph{{IEEE} Trans. Audio, Speech,
  Lang. Process.}, vol.~17, no.~1, pp. 107--116, 2009.

\bibitem{Koyama:IEEE_J_ASLP2013}
S.~Koyama, K.~Furuya, Y.~Hiwasaki, and Y.~Haneda, ``Analytical approach to wave
  field reconstruction filtering in spatio-temporal frequency domain,''
  \emph{{IEEE} Trans. Audio, Speech, Lang. Process.}, vol.~21, no.~4, pp.
  685--696, 2013.

\bibitem{Koyama:JASA_J_2016}
S.~Koyama, K.~Furuya, K.~Wakayama, S.~Shimauchi, and H.~Saruwatari,
  ``Analytical approach to transforming filter design for sound field recording
  and reproduction using circular arrays with a spherical baffle,'' \emph{J.
  Acoust. Soc. Amer.}, vol. 139, no.~3, pp. 1024--1036, 2016.

\bibitem{Miyoshi:IEEE_J_ASSP_1988}
M.~Miyoshi and Y.~Kaneda, ``Inverse filtering of room acoustics,'' \emph{{IEEE}
  Trans. Acoust., Speech, Signal Process.}, vol.~36, no.~2, pp. 145--152, 1988.

\bibitem{Kirkeby:JASA_J_1996}
O.~Kirkeby, P.~A. Nelson, F.~O. Bustamante, and H.~Hamada, ``Local sound field
  reproduction using digital signal processing,'' \emph{J. Acoust. Soc. Amer.},
  vol. 100, no.~3, pp. 1584--1593, 1996.

\bibitem{Daniel:AES114conv}
J.~Daniel, S.~Moureau, and R.~Nicol, ``Further investigations of high-order
  ambisonics and wavefield synthesis for holophonic sound imaging,'' in
  \emph{Proc. 114th {AES} Conv.}, Amsterdam, Netherlands, 2003.

\bibitem{Betlehem:JASA_J_2005}
T.~Betlehem and T.~D. Abhayapala, ``Theory and design of sound field
  reproduction in reverberant environment,'' \emph{J. Acoust. Soc. Amer.}, vol.
  117, no.~4, pp. 2100--2111, 2005.

\bibitem{Ueno:IEEE_ACM_J_ASLP2019}
N.~Ueno, S.~Koyama, and H.~Saruwatari, ``Three-dimensional sound field
  reproduction based on weighted mode-matching method,'' \emph{{IEEE/ACM}
  Trans. Audio, Speech, Lang. Process.}, vol.~27, no.~12, pp. 1852--1867, 2019.

\bibitem{Meyer:ICASSP_2002}
J.~Meyer and G.~Elko, ``A highly scalable spherical microphone array based on
  an orthogonal decomposition of the soundfield,'' in \emph{Proc. {IEEE} Int.
  Conf. Acoust., Speech, Signal Process. ({ICASSP})}, 2002, pp. II--1781--1784.

\bibitem{Abhayapala:ICASSP_2002}
T.~D. Abhayapala and D.~B. Ward, ``Theory and design of high order sound field
  microphones using spherical microphone array,'' in \emph{Proc. {IEEE} Int.
  Conf. Acoust., Speech, Signal Process. ({ICASSP})}, 2002, pp. II--1949--1952.

\bibitem{Park:JASA_J_2005}
M.~Park and B.~Rafaely, ``Sound-field analysis by plane-wave decomposition
  using spherical microphone array,'' \emph{J. Acoust. Soc. Amer.}, vol. 118,
  no.~5, pp. 3094--3103, 2005.

\bibitem{Rafaely:IEEE_J_ASLP_2005}
B.~Rafaely, ``Analysis and design of spherical microphone arrays,''
  \emph{{IEEE} Trans. Audio, Speech, Lang. Process.}, vol.~13, no.~1, pp.
  135--143, 2005.

\bibitem{Ueno:IEEE_SPL2018}
N.~Ueno, S.~Koyama, and H.~Saruwatari, ``Sound field recording using
  distributed microphones based on harmonic analysis of infinite order,''
  \emph{{IEEE} Signal Process. Lett.}, vol.~25, no.~1, pp. 135--139, 2018.

\bibitem{Ueno:IEEE_J_SP_2021}
------, ``Directionally weighted wave field estimation exploiting prior
  information on source direction,'' \emph{{IEEE} Trans. Signal Process.},
  vol.~69, pp. 2383--2395, 2021.

\bibitem{Koyama:WASPAA2021}
S.~Koyama, T.~Nishida, K.~Kimura, T.~Abe, N.~Ueno, and J.~Brunnstr\"{o}m,
  ``{MeshRIR}: A dataset of room impulse responses on meshed grid points for
  evaluating sound field analysis and synthesis methods,'' in \emph{Proc.
  {IEEE} Int. Workshop Appl. Signal Process. Audio Acoust. ({WASPAA})}, Oct.
  2021, (to appear).

\bibitem{Koyama:IEEE_ACM_J_ASLP2020}
S.~Koyama, G.~Chardon, and L.~Daudet, ``Optimizing source and sensor placement
  for sound field control: An overview,'' \emph{{IEEE/ACM} Trans. Audio,
  Speech, Lang. Process.}, vol.~28, pp. 686--714, 2020.

\bibitem{Martin:MultScat}
P.~A. Martin, \emph{Multiple Scattering: Interaction of Time-Harmonic Waves
  with N Obstacles}.\hskip 1em plus 0.5em minus 0.4em\relax New York: Cambridge
  University Press, 2006.

\bibitem{Suzuki:JASA1995}
Y.~Suzuki, F.~Asano, H.-Y. Kim, and T.~Sone, ``An optimal computer-generated
  pulse signal suitable for the measurement of very long impulse responses,''
  \emph{J. Acoust. Soc. Amer.}, vol.~97, no.~2, pp. 1119--1123, 1995.

\end{thebibliography}
%\begin{thebibliography}{00}
%\bibitem{b1} G. Eason, B. Noble, and I. N. Sneddon, ``On certain integrals of Lipschitz-Hankel type involving products of Bessel functions,'' Phil. Trans. Roy. Soc. London, vol. A247, pp. 529--551, April 1955.
%\bibitem{b2} J. Clerk Maxwell, A Treatise on Electricity and Magnetism, 3rd ed., vol. 2. Oxford: Clarendon, 1892, pp.68--73.
%\bibitem{b3} I. S. Jacobs and C. P. Bean, ``Fine particles, thin films and exchange anisotropy,'' in Magnetism, vol. III, G. T. Rado and H. Suhl, Eds. New York: Academic, 1963, pp. 271--350.
%\bibitem{b4} K. Elissa, ``Title of paper if known,'' unpublished.
%\bibitem{b5} R. Nicole, ``Title of paper with only first word capitalized,'' J. Name Stand. Abbrev., in press.
%\bibitem{b6} Y. Yorozu, M. Hirano, K. Oka, and Y. Tagawa, ``Electron spectroscopy studies on magneto-optical media and plastic substrate interface,'' IEEE Transl. J. Magn. Japan, vol. 2, pp. 740--741, August 1987 [Digests 9th Annual Conf. Magnetics Japan, p. 301, 1982].
%\bibitem{b7} M. Young, The Technical Writer's Handbook. Mill Valley, CA: University Science, 1989.
%\end{thebibliography}

\end{document}